\documentclass[%
 amsmath,amssymb,
 aps,
 twocolumn,
]{revtex4-2}

\usepackage{graphicx}
\usepackage{dcolumn}
\usepackage{bm}
\usepackage{hyperref}

\begin{document}

\preprint{APS/123-QED}

\title{Dynamical active particles in the overdamped limit}

\author{Diego M Fieguth }

\affiliation{%
State reaserch center OPTIMAS and Fachbereich Physik, Rheinland-Pf\"alzische Technische Universit\"at Kaiserslautern-Landau,D-67663 Kaiserslautern, Germany
}%

\date{\today}

\begin{abstract}
Mobile microscopic bodies, such as motile cells, can be modelled phenomenologically as ``active particles'' which can move against external forces by depleting an internal energy depot. The microscopic mechanisms underlying such ``active'' behavior must ultimately obey fundamental physics: energy depots must actually consist of dynamical degrees of freedom, such as chemical reaction coordinates, which in some way couple to the particle's motional degrees of freedom. As a step towards connecting phenomenological models with microscopic dynamical mechanisms, recent papers have studied the minimalistic dynamical mechanism of a ``dynamical active particle'', and shown how nonlinear couplings can allow steady energy transfer from depot to motion, even in the presence of weak dissipation. Most real active particles move through viscous environments, however, and are strongly damped. Here we therefore generalize the dynamical active particle into the overdamped regime. We find that its mechanism still operates, and in particular allows the overdamped active particle to travel just as far against friction as the undamped model, by moving at a slower average speed. Our results suggest that active particle phenomenology can indeed be consistent with comprehensible dynamical mechanisms, even in strongly dissipative environments.
\end{abstract}

\maketitle

\section{Introduction}
Our macroscopic world is filled with energy sources that make things move. From planes to toy cars, objects have internal depots of energy that get depleted when moving against friction or other forces. In the microscopic world of biology and bio-physics, objects, such as self propelled particles \cite{Howse2007,Paxton2004} or motile cells \cite{Bodker2010,Friedrich2007,Selmeczi2008} for example, also move against their aqueous, frictional environment, but their inner workings are much less understood. Microscopic moving objects are often modelled by active particles with an internal energy depot\cite{Norden2002,Denisov2002,Cilla2001,Barnhart2010,Romanczuk2011,Kumar2008}.
In recent years active particles have gained much attention, but the models are phenomenological. The enormous question of exactly how internal energy can power active motion at the microscopic level remains open.

Recent papers have shown how a minimal Hamiltonian realisation of such active particles can work, with \cite{DissipativeDaemon} and without \cite{Daemon} weak dissipation. While not intended as precise descriptions of any real active particles, these models have shown that secular energy transfer\cite{Daemon,QDaemon} across large frequency gaps\cite{Henzler2007} can be achieved through nonlinear dynamical couplings, as long as certain non-trivial conditions are obeyed. The strict Hamiltonian framework was in fact stretched in \cite{DissipativeDaemon} by including weak damping, in the standard form of a viscous term in the equation of motion for the particle momentum. A more realistic regime for active particles in aqueous environments, however, is the overdamped Eulerian regime in which viscosity dominates inertia completely, to the point where Newtonian momentum is dynamically irrelevant.

In the present work we take two steps towards a more realistic description: we re-examine the active particle of \cite{DissipativeDaemon,Daemon} in the overdamped, Eulerian limit and we draw connections with established frameworks that give rise to similar equations \cite{aubry2001analytic,Kopidakis2001}.  We find that the nonlinear resonance that enables the energy transfer from depot into motion is \emph{not} a fragile effect, but instead robustly adapts to viscosity by driving the active particle more slowly. 

Our results demonstrate that dynamical mechanisms with very few moving parts can be sufficient to make microscopic bodies behave like powered vehicles, while still being simple enough to operate well under strong dissipation.

\subsection{Organisation}
The paper is structured as follows. In Section \ref{sec:2} we will give a brief description of the mechanism that allows for dynamical active particle (DAP) behaviour and discuss possible implementations. In section \ref{sec:3} we then present the overdamped equations and show numerical solutions for two conceptually different parameter regimes. From these numerical simulations we can infer appropriate approximation methods which we apply in \ref{sec:4}. There we apply the classical averaging theorem to the overdamped equations and perform a fixed point analysis. This leads to our main result: the dynamical active particle performs work as efficiently as in the underdamped case.  In section \ref{sec:5} we add Brownian noise to the problem and obtain approximations to the resulting stochastic differential equation by using previous results. We give a simple expression for the mean square displacement which we compare to ensemble averages obtained by numerical realisations of the stochastic differential equation. We end with a brief conclusion in \ref{sec:6}.

\section{Setup of the active particle Hamiltonian}\label{sec:2}
In this section we will present the underlying Hamiltonian model that admits downconversion in the underdamped case. The key feature is a non-linear resonance of the particular form studied in detail by Chirikov\cite{Chirikov}.

The model we will look at was first presented without damping in \cite{Daemon}, where the Hamiltonian was introduced and its canonical equations of motion were studied. In \cite{DissipativeDaemon} the system was extended to include dissipative effects. This is done within the Hamiltonian framework by adding a Hamiltonian environment and interaction. From this it is standard procedure to derive effective equations that are not Hamiltonian themselves, but are compatible with Hamiltonian mechanics \cite{StochasticEnergetics}. 

The Hamiltonian loses its role as sole generator of time evolution, as it only generates part of it, but it still allows us to describe the microscopic mechanism of a possible active particle in a way that is consistent with fundamental classical physics, rather than being purely phenomenological. By adding strong damping we can now evaluate whether such a simple microscopic mechanism can operate in the overdamped regime.

\subsection{The Hamiltonian Model}
The Hamiltonian model is split into three parts: a mass, the energy depot and a small coupling. With the pairs of canonical coordinates $(q,p)$ and $(\alpha,I)$ we have \begin{align}
    H_\text{M}(q,p) &= \frac{p^2}{2M} + fq,\label{eq:HM} \\ 
    H_\text{D}(\alpha,I) &= \omega I\;\label{eq:HD}\\
    H_\text{C}(q,\alpha,I) &= -\omega\sqrt{I_0^2-I^2} \cos(kq-\alpha)\;\label{eq:HC}.
\end{align} The Hamiltonian of a mass $H_\mathrm{M}$ includes the usual kinetic term with momentum $p$ and a linear potential that results in an opposing force $-f$. This external force is an idealisation which could easily be relaxed to include any conservative force that does not vary too rapidly with position.

$H_\mathrm{D}$ uses a generic action variable to represent the energy depot. In \cite{Daemon} this action variable is the difference of actions of two harmonic oscillators of different frequencies. A more plausible implementation of such an energy depot are discrete breathers \cite{Kopidakis2001,aubry2001analytic}, which are often used in models for protein structures \cite{piazza2011breather,luccioli2011discrete}. While more complicated than harmonic oscillators it is still possible to find a collective action variable to describe them. In \cite{aubry2001analytic} it was shown how to transfer energy between two such systems efficiently and irreversibly if a resonance condition is met. We will show that we do not need such a strict resonance condition when coupling the system to an overdamped degree of freedom.

The final ingredient is the coupling between the mass and the energy depot, $\varepsilon H_\mathrm{C}$. This coupling is small to justify the separation into two subsystems. This smallness is indicated by the parameter $\varepsilon\ll1$. The important feature of this coupling in \cite{Daemon,DissipativeDaemon} in the underdamped case is that it has to give rise to a non-linear resonance \cite{Chirikov}. This resonance exists in some region of phase space and thus can be reached dynamically, opposed to a resonance that occurs in parameter space, where parameters need to be fine tuned \cite{aubry2001analytic}. With this non-linear resonance it is possible to accomplish the task of transferring energy from fast degrees of freedom (the high frequency energy depot) to a slow degree of freedom (the mass), without the need for external control. This periodic coupling can be seen as the lowest order expansion (beyond being constant) of a small coupling \cite{Kopidakis2001}.

In this paper we will use the model studied before in \cite{DissipativeDaemon} and \cite{Daemon}, in which the canonical variables $(\alpha,I)$ represent two harmonic oscillators after an additional adiabatic elimination. 
The ratio $\frac{\omega}{k}=:v_\mathrm{c}$ determines the resonant speed of the Chirikov resonance. The constant $I_0$ determines how much energy can be transferred. 

The exposure to a frictional environment is achieved by simply adding a linear friction term $-\frac{\gamma}{M}p$ to the equation of motion for the momentum of the weight $p$.
This leads us to the following equations for the dynamics of the system \begin{align} \dot{q} & =\frac{p}{M} \label{eq:ud:q},\\ \dot{p} & =-f-k \varepsilon \omega\sqrt{I_0^2-I^2} \sin (k q-\alpha)-\frac{\gamma}{M} p, \\ \dot{\alpha} & =\omega+\varepsilon \frac{\omega I}{\sqrt{I_0^2-I^2}} \cos (k q-\alpha), \\ \dot{I} & =\varepsilon \omega \sqrt{I_0^2-I^2} \sin (k q-\alpha)\label{eq:ud:I}.\end{align}

The canonical momentum $I$ is proportional to the depot energy $\omega I$. Also in \eqref{eq:ud:I} we see that the depots values are between $-I_0<I<I_0$.

\begin{figure}
    \centering
    \includegraphics[width=0.5\textwidth]{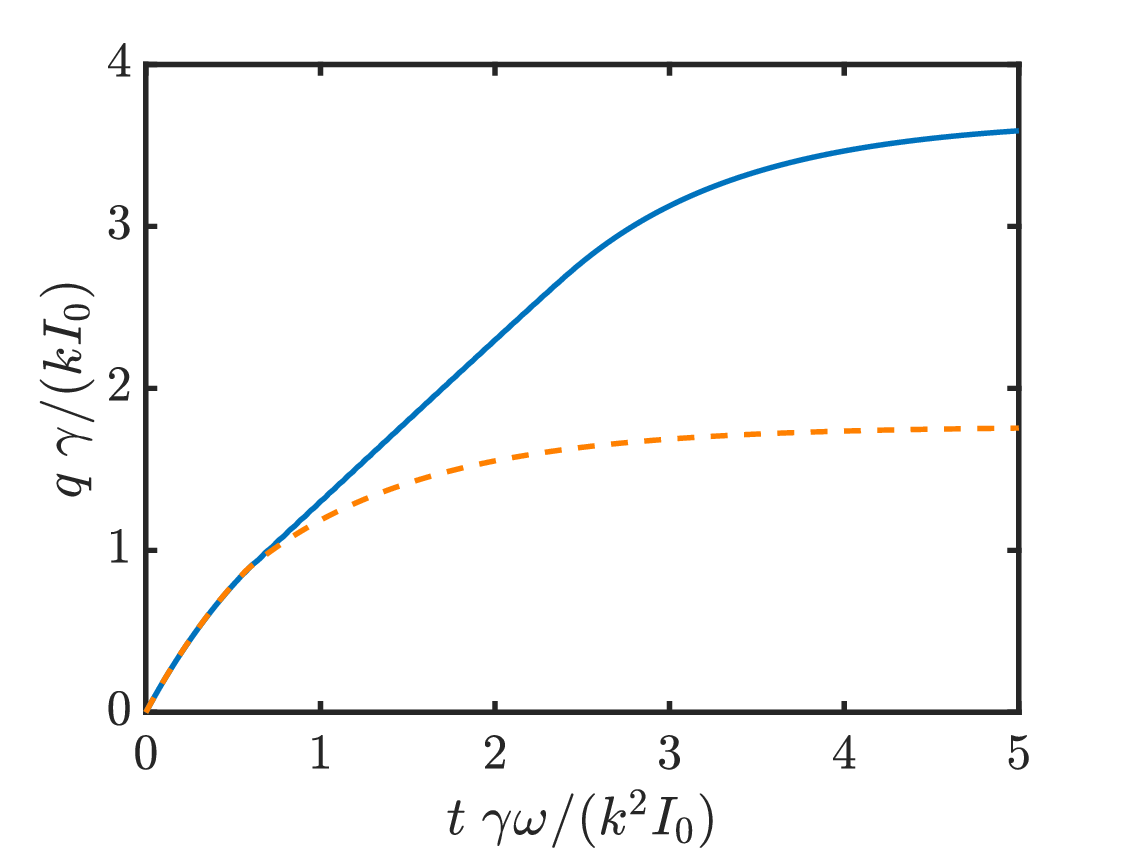}
    \caption{Scaled position over scaled time for numerical solutions to the equations of the underdamped DAP \eqref{eq:ud:q}-\eqref{eq:ud:I}, with two distinct behaviours due to different initial conditions $\alpha(0)$.}
    \label{fig:Underdamped Daemon}
\end{figure}

For weak damping $\gamma/M=\mathcal{O}(\varepsilon)$, two kinds of trajectories are possible \cite{DissipativeDaemon,Daemon}; see Figure \ref{fig:Underdamped Daemon}. When reaching the critical speed $v_\mathrm{c}=\omega/k$ the depot energy is either used to keep the speed from changing (blue trajectory) or the two subsystems behave as if decoupled and we observe the usual damping due to friction (dashed orange trajectory). 

In the overdamped case in contrast, only one kind of trajectory exists and all initial conditions with energy in the depot lead to a transfer of this energy.

\section{Overdamped equations}\label{sec:3}
In frictional environments, inertial terms tend to zero for a small mass. There exists an appropriate limit to the classical Langevin equation, called the overdamped Langevin equation or sometimes the Smoluchowski-Kramers approximation \cite{Smoluchowski,Kramers1940,Freidlin2004}. These models include Brownian white noise, which we will add in Section \ref{sec:5}. First we will study the behaviour in the zero noise limit.

For small mass $M$ the inertial term $M\ddot{q}=\dot{p}$ tends to zero. In the limit $M\to 0$ we can set $\dot{p}=0$ to get an expression for $p$, which we can then insert into the equation for $\dot{q}$. The overdamped equations are calculated by setting

\begin{equation}
    \dot{p}=0=-f-k\varepsilon\omega\sqrt{I_0^2-I^2}\sin(kq-\alpha)-\frac{\gamma}{M} p
\end{equation}
which yields \begin{equation}
    p=-\frac{Mf}{\gamma}-\frac{\varepsilon \omega M k}{\gamma}\sqrt{I_0^2-I^2}\sin(kq-\alpha) \label{eq:preq}
\end{equation}

and gives us three equations that describe the motion of the system after inserting \eqref{eq:preq} into \eqref{eq:ud:q}, \begin{align}
    \dot{q}&=-\frac{f}{\gamma}-\frac{\varepsilon k \omega}{\gamma}\sqrt{I_0^2-I^2}\sin(kq-\alpha),\label{eq:q}\\
    \dot{\alpha} & =\omega+\varepsilon \frac{\omega I}{\sqrt{I_0^2-I^2}} \cos (k q-\alpha),\label{eq:alpha} \\
    \dot{I} & =\varepsilon \omega \sqrt{I_0^2-I^2} \sin (k q-\alpha)\label{eq:I}.
\end{align}

Since all expressions only depend on the difference of $kq-\alpha$ we can combine the first two equations into one equation for $\varphi:=\alpha-kq$,\begin{equation}
    \dot{\varphi}=\omega+\frac{kf}{\gamma}+\varepsilon\omega\left(\frac{I}{\sqrt{I_0^2-I^2}} \cos (\varphi) -\frac{k^2}{\gamma}\sqrt{I_0^2-I^2}\sin(\varphi) \right)\label{eq:phi2d}
\end{equation}
which results in the depot energy obeying\begin{equation}
\dot{I}=-\varepsilon\omega\sqrt{I_0^2-I^2}\sin(\varphi). \label{eq:I2d}
\end{equation}

\begin{figure}[ht]
    \centering
    \includegraphics[width=0.5\textwidth]{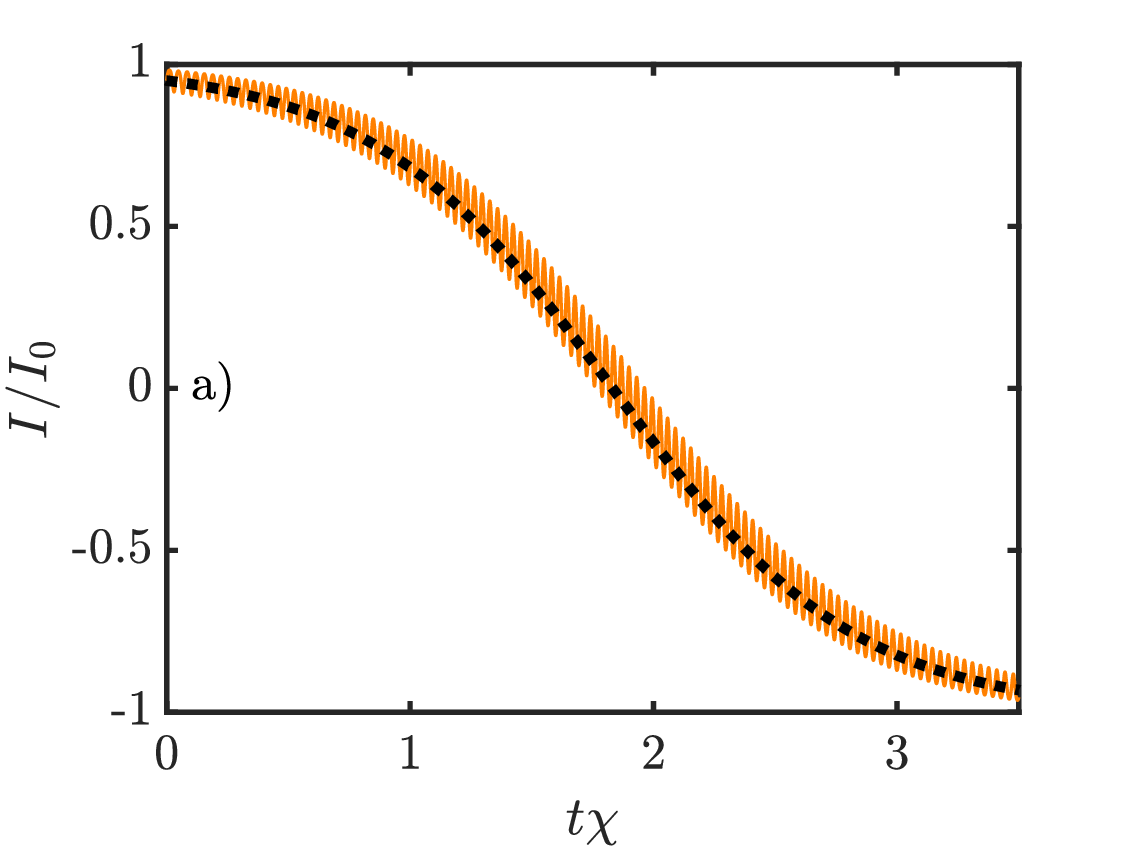}
    \includegraphics[width=.5\textwidth]{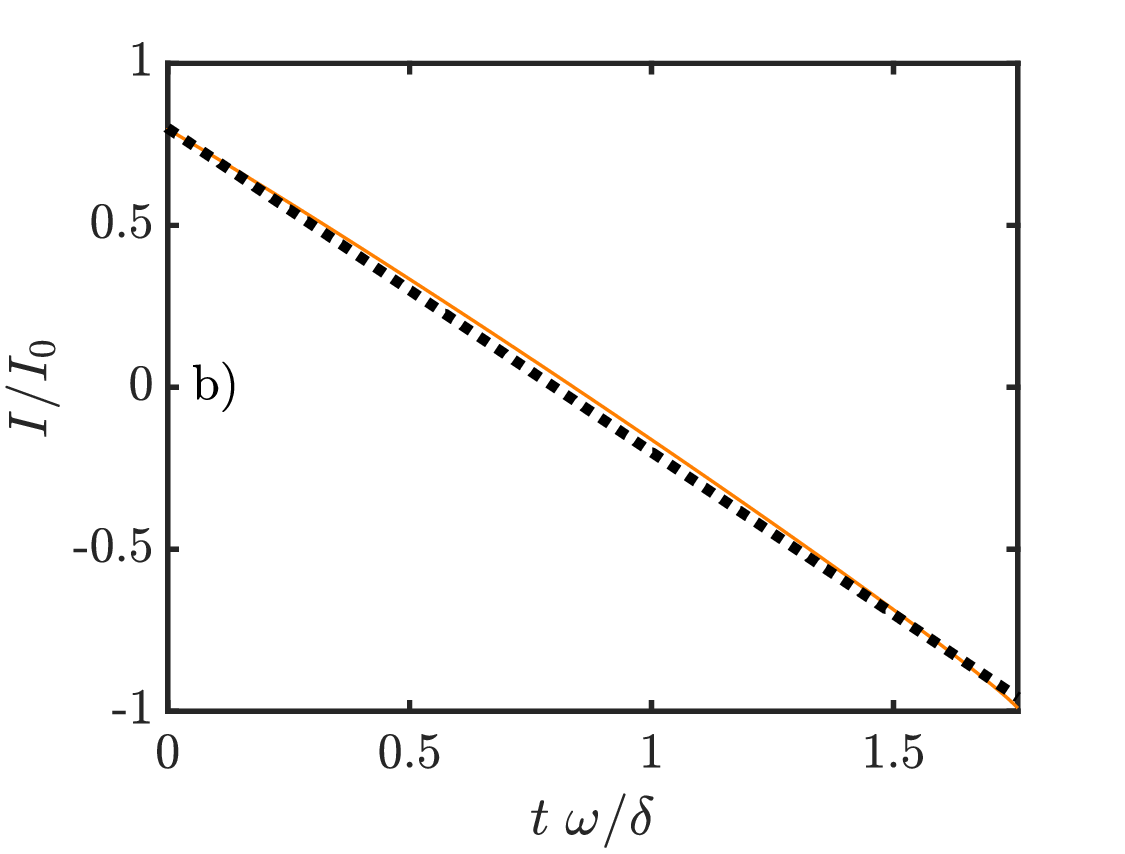}
    \caption{Solution of the depot energy of the dynamical active particle in the two qualitatively different regimes. In a): Solution of rescaled $I$ over rescaled time for the two different frictional regimes explained in the text. gained by numerically integrating \eqref{eq:phi2d},\eqref{eq:I2d} (solid orange line) with initial conditions $\varphi(0)=-\pi/2,I(0)=0.95$. In a) we can also see the averaged solution \eqref{eq:Iav} for $I(0)=0.95$ (thick black line) for $\omega=k=\gamma=1$, $f=0$ and $\varepsilon=0.1$. The parameter $\chi$ is defined in \eqref{eq:chi}. These parameters where chosen to be illustrative and show the effect of the different timescales.
    In b): Solution of rescaled $I$ over rescaled time, gained by numerically integrating \eqref{eq:phi2d},\eqref{eq:I2d} (solid orange line) with initial conditions $\varphi(0)=2,I(0)=0.8$. In b) we can also see the approximate solution \eqref{eq:linearI} for $I(0)=0.8$ (dashed black line) for $\omega=\gamma=M=1$, $f=0$, $k=25$ and $\varepsilon=0.1$. The parameter $d$ is defined in \eqref{eq:compareED}.}
    \label{fig:Ievo}
\end{figure}

The time evolution of $I$ obtained by numerical integration of \eqref{eq:phi2d},\eqref{eq:I2d} can be seen as the orange line in Figure \ref{fig:Ievo} for illustrative parameters. The black line is the averaged solution discussed in Section \ref{sec:4}. We can clearly see that the depot is depleting, from which we expect movement since energy can only be lost to friction. The simple dependence of $q$ on the depot energy can be seen in \eqref{eq:q} in the following way: Noticing that the second term is just $-\frac{k}{\gamma} \dot{I}$, we can integrate \eqref{eq:q} which gives us \begin{equation}
    q(t)=q(0)-\frac{f}{\gamma}t-\frac{k}{\gamma}(I(t)-I(0)). \label{eq:unaveragedQ}
\end{equation}
This confirms that decreasing $I$ means increasing $q$.
From equation \eqref{eq:unaveragedQ} we also see that the dimensionless parameter\begin{equation}
    d=\frac{k^2I_0}{\gamma}
\end{equation} is important for the system since the travel distance is of order $d/k$.

In Figure \ref{fig:Ievo} we can see numerical solutions of the overdamped equations for the DAP for two different values of $d$ that give rise to qualitatively different evolutions. 
In Figure \ref{fig:Ievo} a) we have $\varepsilon d\ll1$. We can see small oscillations due to the fast rotating phase $\varphi$. For problems involving such a fast angle $\varphi$ we can average the evolution for $I$. The averaged evolution is sufficient since \eqref{eq:unaveragedQ} relates the depleted depot energy to displacement. In Figure \ref{fig:Ievo} b) we have $\varepsilon d\gg1$. There we see a different behaviour. In this regime the depot energy depletes linearly, without oscillations. These two regimes of different values of $\varepsilon d$ will be investigated in the next section.

\section{Analytical description of the behaviour}\label{sec:4}
In the last section we derived the overdamped equations of the Hamiltonian system from Section \ref{sec:2}. We reduced the set of equations of motion to two, namely the coordinates $(\varphi,I)$. So the full dynamics of the system will be described by these two variables. The quantity we are interested in however is the (generalised) position $q$. We want to know how far the DAP will move in the frictional environment. This highlights the importance of \eqref{eq:unaveragedQ}, which relates $q$ and $I$. This means any analytical approximation of the depot variable $I$ will give us an approximation of the displacement.

Here we will make suitable approximations in two complementary parameter regimes of small and large $\varepsilon d$ shown in Figure \ref{fig:Ievo}.

First we find a simplified (averaged) description of the dynamics in the $(\varphi,I)$-system, yielding a simple time evolution for $I$, which we then can use to determine the displacement using \eqref{eq:unaveragedQ}. As we will see, averaging requires $\varepsilon d$ to be small. The case of large $\varepsilon d$ is investigated separately at a later point in this section.
\subsection{Application of the averaging Theorem}
Here we will use standard averaging arguments due to the fact that $\omega$ is the fastest time scale in the system.
From the time evolution in Figure \ref{fig:Ievo} a) we see that before $I$ is changing significantly there are many small oscillations resulting from the fast evolution of $\varphi$. We will now average the equation for $I$, obtaining an average value $\bar{I}$, and derive an expression for the thick black line in Figure \ref{fig:Ievo} using the fact that $\varepsilon$ is small. For brevity we will also assume that the external force is $f=0$.
We first look at the integral of $\dot{I}$ over a short time $\Delta t\gg\omega^{-1}$\begin{equation}
    \Delta I=\int\limits_0^{\Delta t}\!\mathrm{d}t \dot{I}=\int\limits_0^{\Delta\varphi}\!\mathrm{d}\varphi \frac{\dot{I}}{\dot{\varphi}}.\label{eq:deltaI0}
\end{equation}

Since $I$ changes on a much slower scale than $\varphi$, we can approximate the integral by\begin{equation}
    \Delta I\approx\frac{\omega \Delta t}{2\pi}\int\limits_0^{2\pi}\!\mathrm{d}\varphi \frac{\dot{I}}{\dot{\varphi}}.\label{eq:deltaI1}
\end{equation}  Colloquially speaking, the prefactor in \eqref{eq:deltaI1} determines how often the integral in \eqref{eq:deltaI1} fits into the integral expression in \eqref{eq:deltaI0}. Also $\dot{I}$ and $\dot{\varphi}$ should be understood as functions of $\varphi$ and the current average value $\bar{I}$. 
Inserting \eqref{eq:phi2d} and \eqref{eq:I2d}, we are now left with the coarse grained change of $I$ given by\begin{align}
    \dot{\overline{I}}&:=\frac{\Delta I}{\Delta t}=-\frac{\varepsilon\omega}{2\pi}\sqrt{I_0^2-\bar{I}^2} \times \nonumber \\ &\int\limits_0^{2\pi}\mathrm{d}\varphi \sin(\varphi) \left(1+\varepsilon\left(\frac{\bar{I} \cos (\varphi)}{\sqrt{I_0^2-\bar{I}^2}} -\frac{k^2}{\gamma}\sqrt{I_0^2-\bar{I}^2}\sin(\varphi) \right)\right)^{-1}. \label{eq:firstIav}
\end{align}

It is easy to see that the zeroth order in $\varepsilon$ of the integrand vanishes since we integrate $\sin(\varphi)$ over one period. The first order Taylor expansion in $\varepsilon$ of the integrand in \eqref{eq:firstIav} yields \begin{equation}
    -\int\limits_0^{2\pi}\mathrm{d}\varphi \sin(\varphi)\left(\frac{\bar{I}}{\sqrt{I_0^2-\bar{I}^2}} \cos (\varphi) -\frac{k^2}{\gamma}\sqrt{I_0^2-\bar{I}^2}\sin(\varphi) \right),\label{eq:expansion}
\end{equation} which is a viable approximation as long as the dimensionless travel distance fulfils\begin{equation}
    d\equiv\frac{k^2I_0}{\gamma}\ll\frac{1}{\varepsilon}. \label{eq:compareED}
\end{equation} The case of large $\varepsilon d$ will be considered separately.

The product $\sin(\varphi)\cos(\varphi)$ also vanishes after integrating and we are left with a non-vanishing integral over $\sin^2\varphi$.
This leads us to the averaged evolution \begin{equation}
\dot{\overline{I}}=-\frac{\varepsilon^2k^2\omega}{2\gamma}(I_0^2-\bar{I}^2)+\mathcal{O}(\varepsilon^3) \label{eq:averageI}
\end{equation}

which, after neglecting higher order terms, has the simple solution \begin{equation}
    \overline{I}(t)=-I_0\tanh \left(\chi (t-t_0)\right), \label{eq:Iav}
\end{equation}
where \begin{equation}\chi=\frac{\varepsilon^2k^2\omega I_0}{2\gamma }\label{eq:chi}\end{equation} and $t_0=\chi^{-1}\mathrm{arctanh}(\overline{I}(0)/I_0) $. 
The averaged solution \eqref{eq:Iav} can be seen in Figure \ref{fig:Ievo} as the thick black line. Even though the numerical evolution was done for a rather large $\varepsilon=0.1$, the averaged solution describes the secular depletion of the depot energy well as long as $d$ is not too large (see \eqref{eq:compareED}).
\begin{figure}[p]
    \centering
    \includegraphics[width=.5\textwidth]{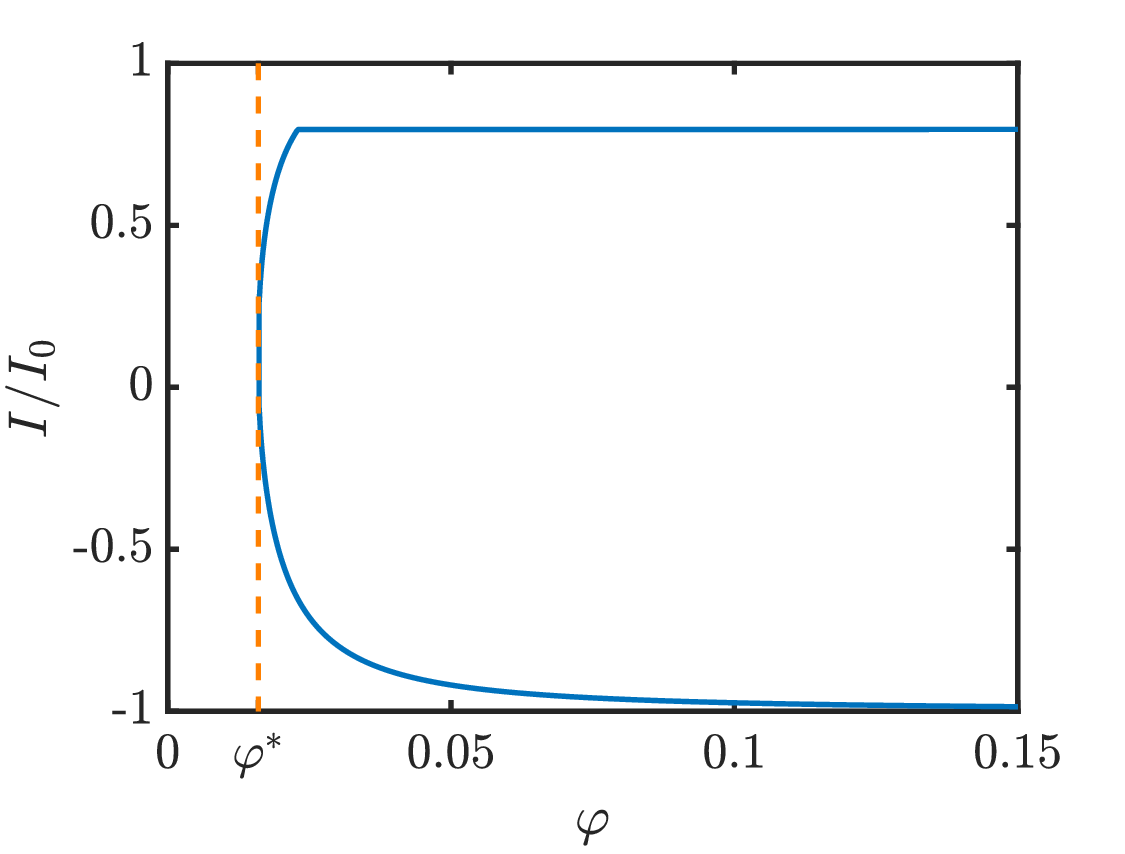}
    \caption{Numerical solution to \eqref{eq:q}-\eqref{eq:I}. The trajectory in phase space (blue) can be seen with the limiting value $\varphi^\ast$ (dashed orange). This is the corresponding phase space evolution to Figure \ref{fig:Ievo} b).}
    \label{fig:HighDelta}
\end{figure}

\subsection{Fixed point analysis}\label{sec:4b}
The DAP enters a qualitatively different regime when \eqref{eq:compareED} does not hold. There is a critical value for which $\varphi$ has a fixed point $\varphi^\ast$, namely if \eqref{eq:phi2d} is zero. If the value of $I$ is not too close to the extremal values $\pm I_0$, we can neglect the first order of $\varepsilon$ in \eqref{eq:phi2d} and set it to zero to obtain \begin{equation}
\varepsilon d \sqrt{1-\left(I/I_0\right)^2}\sin(\varphi^\ast)=1.
\end{equation}
This equation has real solutions if \begin{equation}
    \varepsilon d>1.
\end{equation} Then there is a fixed point in a region with \begin{equation} 
|I|<I_0\sqrt{1-(\varepsilon d)^{-2}}. \label{eq:conditionI}
\end{equation}
This yields a stationary phase value $\dot{\varphi}=0$ around $I=0$.
In regions where condition \eqref{eq:conditionI} holds, the depot energy changes like $\dot{I}=-I_0\omega/d$ with an error of order $\varepsilon \omega\sqrt{I_0^2-I^2}$. Around $I(0)=0$, this leads to a solution \begin{equation}
    I^{(0)}(t)=-I_0\frac{\omega}{d} t \label{eq:linearI}
\end{equation}
in zeroth order in $\varepsilon$. This zeroth order solution can be seen as the dashed black line in Figure \ref{fig:Ievo} b). The deviation in Figure \ref{fig:Ievo} b) can be explained by looking at higher orders. The first order in $\varepsilon$, $I^{(1)}$, for example can be obtained by inserting the zeroth order solutions $(\varphi^{(0)}=\varphi^\ast,I^{(0)})$ into the equation of motion, \begin{equation}
    \dot{I}^{(1)}=-\varepsilon\omega\sqrt{I_0^2-I^{(0)}(t)^2}\sin(\varphi^\ast)=-\varepsilon I_0\omega\sqrt{1-\frac{\omega^2}{d^2}t^2}.
\end{equation}

The behaviour of the phase $\varphi$ in this regime can be seen in Figure \ref{fig:HighDelta}. The trajectory of the numerical solutions is shown in the $(\varphi,I)$-phase-space in blue. We clearly see that it approaches the value $\varphi^\ast=\mathrm{arcsin}((\varepsilon d)^{-1})$ around $I=0$. Near this fixed value, the depot energy changes. The time evolution of the depot energy corresponding to Figure \ref{fig:HighDelta} was shown in Figure \ref{fig:Ievo} b). There we can confirm the depleting depot energy, which in return means increasing position $q(t)$ (see \eqref{eq:unaveragedQ}).\newline

This means we can confirm the depleting depot in two opposing regimes, of small and large $\varepsilon d$.
Now that we know the scales involved in the evolution of the energy in the depot for different parameter regimes, we can discuss the efficiency.

\subsection{Efficiency} \label{sec:4 Efficiency}
We will now compare the DAP to an energy transfer by a kick and examine how far the particle can move. What we call efficiency is a measure of displacement per used energy. The kick will accelerate the particle instantaneously to a velocity such that the kinetic energy has a value $E_0$. 
An overdamped particle follows Newton's equation \begin{equation}
    M\ddot{q}=-\gamma \dot{q}.
\end{equation}
Integrating this equation yields \begin{equation}
    \dot{q}=-\frac{\gamma}{M}(q+c),\label{eq:free overdamped}
\end{equation} where $c$ is an integration constant. We set $c=0$ since it only constitutes a shift in position and we care about relative changes. Equation \eqref{eq:free overdamped} has the solution \begin{equation}
    q(t)=L\exp\left(-\frac{\gamma}{M}t\right),
\end{equation} where $L$ is the distance travelled against friction. The initial energy is the kinetic energy since no potentials are involved, \begin{equation}
    E_0=\frac{M}{2}\dot{q}(0)^2=\frac{M}{2}\left(\frac{L\gamma}{M} \right)^2=\frac{L^2\gamma}{2M}.
\end{equation} The displacement as a function of the input energy is then \begin{equation}
    L_\mathrm{kick}(E_0)=\sqrt{\frac{2ME_0}{\gamma^2}}. \label{eq:displacement1}
\end{equation} We see that the distance travelled vanishes in the overdamped limit $M\to0$, so the efficiency tends to zero.

\subsubsection{Efficiency of the DAP}
The efficiency for both $\varepsilon d$-regimes is the same. From \eqref{eq:unaveragedQ} it is clear that the maximum distance travelled is of order \begin{equation}
    L_\mathrm{DAP}=\frac{kI_0}{\gamma}=d/k.
\end{equation}
The kinetic energy scales with $M$ and hence vanishes in the limit $M\to0$. This means that in the overdamped limit all the initial energy is in the depot and is of order \begin{equation}
    E_0=I_0\omega.
\end{equation}
This leads to a relation of displacement and energy, \begin{equation}
    L_\mathrm{DAP}=\frac{k E_0}{\omega\gamma}=\frac{E_0}{v_\mathrm{c}\gamma}.\label{eq:displacement2}
\end{equation}
Comparing the displacements \eqref{eq:displacement1} and \eqref{eq:displacement2}, we see that the DAP is far more effective in using the initial energy $E_0$ to gain displacement since the distance is not affected by the mass $M$. The travelled distance is the initial energy times the displacement per used energy $1/(\gamma v_\mathrm{c})$.

The efficiency is the same for both $d$ regimes; it is however achieved in different ways. In the small $\varepsilon d$-regime the particle moves slower so we might expect it to be more efficient. The motion in this regime consists of many back and forth oscillations with a slowly moving average, so a lot of energy is wasted for non-secular motion.

In the large $\varepsilon d$-regime, the DAP is faster and hence loses more energy per time to the environment. The movement in this case is straight forward and there are no oscillating terms that need to be taken into account. Using the zeroth order approximation of $I$ from equation \eqref{eq:linearI} the velocity is is \begin{equation}
    \dot q=-\frac{k}{\gamma} \dot{I}=\frac{k^2}{k\gamma}\frac{I_0\omega}{d}=\frac{\omega}{k}=v_\mathrm{c}.
\end{equation}
This results in the same velocity that was the resonant velocity in the underdamped case we reviewed in Section \ref{sec:2}.

Even though the mechanisms are quite different, the displacement per used energy is the same for fixed $\gamma v_\mathrm{c}$.

\section{Brownian environment}\label{sec:5}
An active particle does not only need to work against strong frictional forces, but also needs to continue to work in noisy environments. In \cite{DissipativeDaemon} noise was included in numerical simulations for the underdamped case.
In this section we add Brownian white noise and apply our knowledge about the zero noise limit from the previous section. 
\subsection{Noise in the small $\varepsilon d$-regime}
The Brownian motion has a variance given by the Einstein relation $\sigma^2/2=\gamma k_\mathrm{B}T$ \cite{StochasticEnergetics}. Adding noise to our problem yields 
\begin{equation}
    \dot{q}=-\frac{f}{\gamma}-\frac{\varepsilon \omega k}{\gamma}\sqrt{I_0^2-I^2}\sin(kq-\alpha)+\sigma\dot{W}_t=\frac{f}{\gamma}-\frac{k}{\gamma}\dot{I}+\sigma\dot{W}_t \label{eq:q with noise}
\end{equation} where $\dot{W}_t$ is to be interpreted as a Wiener process in the sense of distributions. Equations \eqref{eq:alpha} and \eqref{eq:I} do not change.
Cases of strong and weak noise are trivial, the influence is either overpowering or negligible. In the small $\varepsilon d$-regime the noise is non-trivial if we have $\sigma=\mathcal{O}(\varepsilon)$. It is for this noise intensity that the $\dot{I}$-term and the noise term in \eqref{eq:q with noise} are of the same magnitude over long times; the noise changes on the same time-scale as the averaged change of the depot energy \eqref{eq:Iav}.

Apart from the direct effect of the noise on $q$, it affects $I$ via the angle variable in its equation of motion \eqref{eq:I}.

If we repeat the same averaging procedure as in Section \ref{sec:4}, we see that the noise term in $\bar{I}$ can be neglected since it is of order a higher order in $\varepsilon$ in expansion \eqref{eq:expansion} (keeping in mind that the order of the noise scales with $\sigma^2$).

For a vanishing force $f$, and $\sigma^2=\mathcal{O}(\varepsilon^2)$ it holds that \begin{equation}
    q(t)=q(0)+\frac{k}{\gamma}I(0)+\frac{k}{\gamma}\tanh(\chi(t-t_0))+\sigma W_t +\mathcal{O}(\varepsilon^3).
\end{equation}

Keeping only the lowest orders of $\varepsilon$, this leads to a mean square displacement behaving like \begin{align}
    \langle q^2 \rangle(t)&=\left \langle \left[q(0)+\frac{k}{\gamma}I(0)+\frac{k}{\gamma}\tanh(\chi(t-t_0))+\sigma W_t\right]^2\right\rangle,\\
    &=\left[q(0)+\frac{k}{\gamma}I(0)+\frac{k}{\gamma}\tanh(\chi(t-t_0))\right]^2\\ &\,+2 \sigma\left[q(0)+\frac{k}{\gamma}I(0)+\frac{k}{\gamma}\tanh(\chi(t-t_0))\right]\langle W_t\rangle \nonumber \\ &\qquad+\sigma^2\langle W_t^2\rangle,\\
    &=\left[q(0)+\frac{k}{\gamma}I(0)+\frac{k}{\gamma}\tanh(\chi(t-t_0))\right]^2+\sigma^2t.\label{eq:MDS}
\end{align}

\begin{figure}[ht]
     \centering
     \includegraphics[width=.45\textwidth]{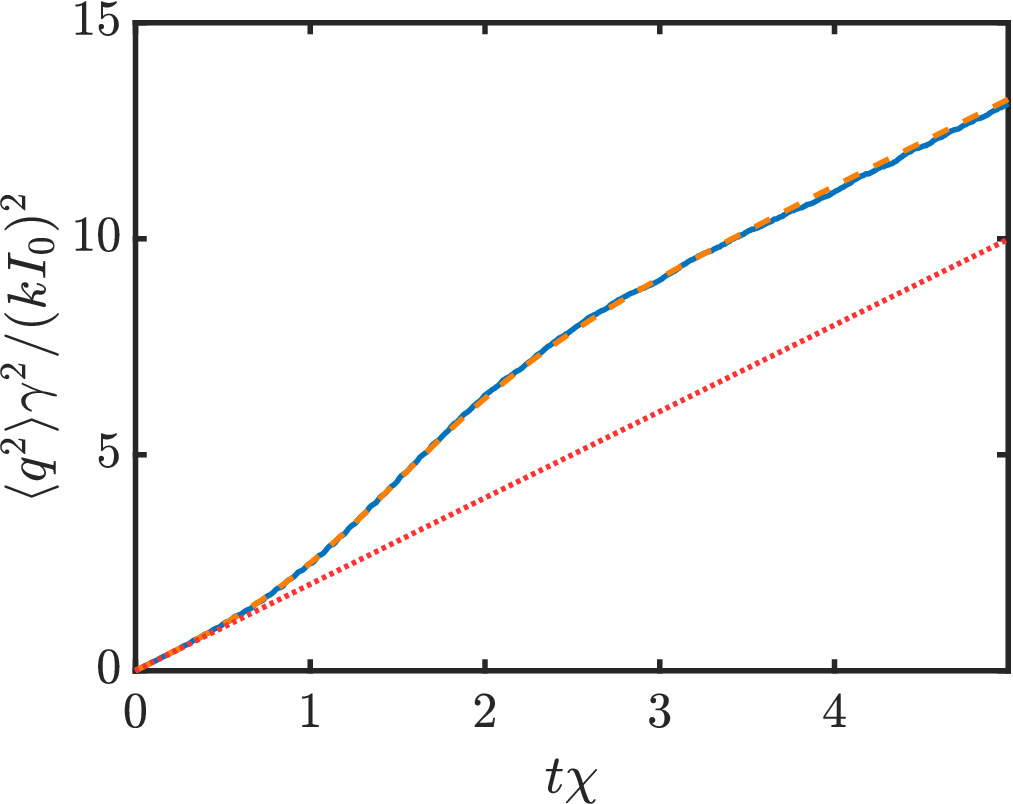}
    \caption{The (scaled) mean squared displacement of $10^4$ sample trajectories (blue) obtained by solving \eqref{eq:q}-\eqref{eq:I} with additional white noise in \eqref{eq:q} plotted over (scaled) time. The dashed orange corresponds to the analytical result of  result of \eqref{eq:MDS}. The red dotted line depicts the diffusive term $\sigma^2t$. The numerical integration was performed for $\omega=\sigma\varepsilon^{-2}=\gamma=k=I_0=1$, $f=0$ and $\varepsilon=0.01$ with initial conditions $I(0)=0.8,q(0)=0,\alpha(0)=-\pi/2$.}
     \label{fig:MDS}
 \end{figure}

This analytical result can be seen in Figure \ref{fig:MDS} as the dashed orange line. The blue line is the mean square displacement over numerically simulated trajectories of the DAP in a Brownian environment. The dotted red trajectory shows the mean square displacement in absence of the energy depot. We see that on top of the diffusion there is a clear displacement that has its origins in the conversion of the depot energy we have seen in the previous section. When the depot is nearly depleted the mean squared displacement continues to grow with the diffusive growth constant.

\subsection{Noise in the large $\varepsilon d$-regime}

For small values of $\varepsilon d$ the case non-trivial noise intensity turned out to relate in a simple way to the zero-noise solutions we derived in Section \ref{sec:4}. For large $\varepsilon d$ this is not the case. Instead of giving an exhaustive statistical analysis, we will start with weak noise and show numerically the onset of new behaviours when the noise intensity increases.

\begin{figure}[ht]
    \centering
    \includegraphics[width=.5\textwidth]{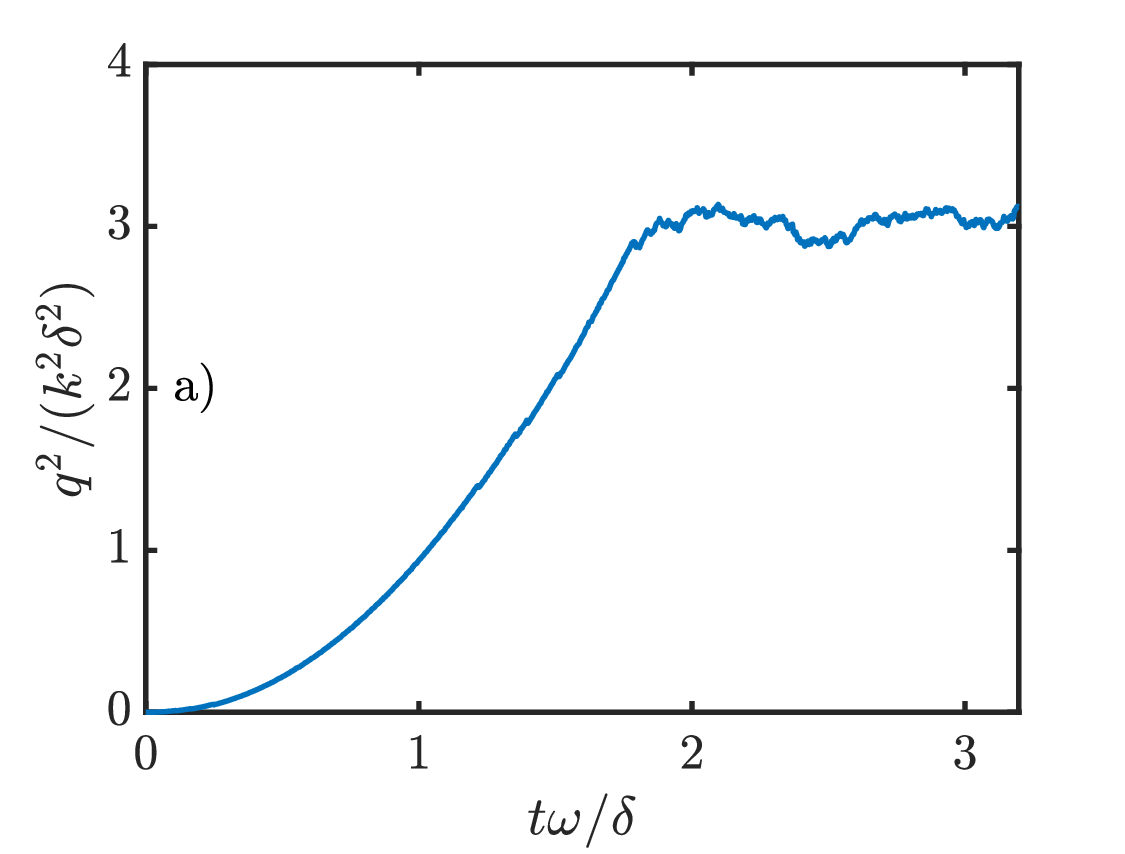}
     
     \includegraphics[width=.5\textwidth]{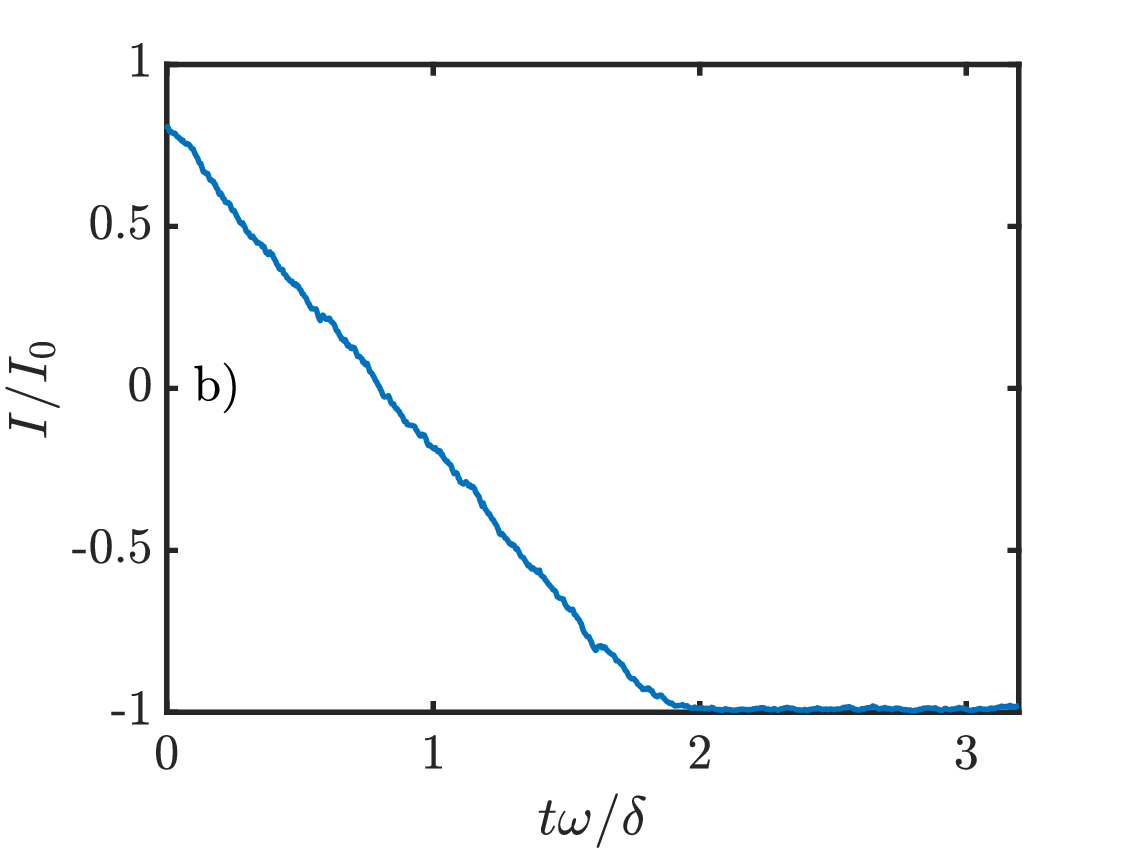}

     \includegraphics[width=.5\textwidth]{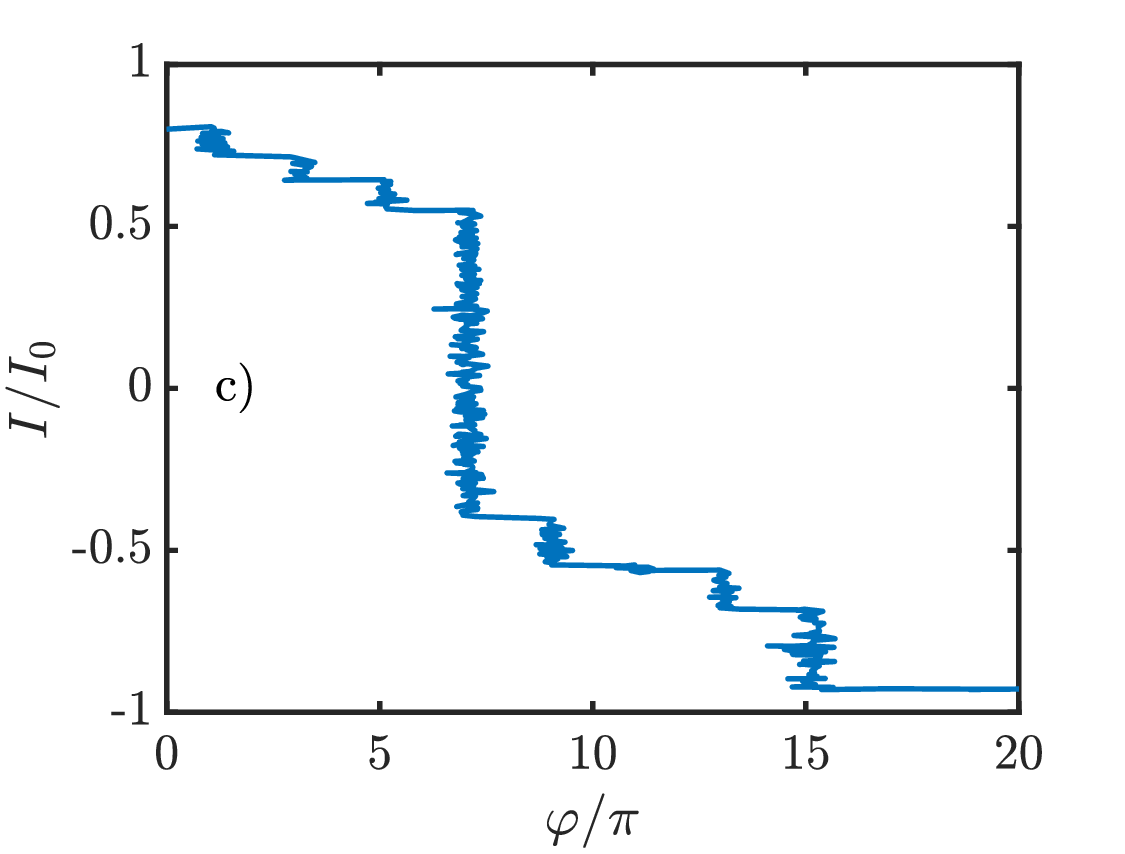}
    \caption{Numerical solution to \eqref{eq:q}-\eqref{eq:I} with added noise $\varepsilon\sigma\dot{W}_t$ in \eqref{eq:q}. In a) we see the time evolution of the position. In b) the time evolution of the depot energy is depicted (blue). In c) the trajectory in the $(\varphi,I)$-space is shown. The numerical integration was performed for $\omega=\gamma=I_0=1$, $f=0$, $k=25$ and $\varepsilon=0.01$. To best show the transient effects we chose $\sigma/\sqrt{\varepsilon}=0.7$.}
    \label{fig:HighDeltaDifferentNoise}
\end{figure}

Change in the large $\varepsilon d$ regime happens on a time of order $\varepsilon^0$ since the particle moves with at the critical speed $v_\mathrm{c}$. We now choose the noise intensity to be of order $\sigma^2=\varepsilon$, so the influence of the noise is small compared to the displacement generated by the DAP. Over a time interval such that $t\omega/d=\mathcal{O}(1)$ the squared displacement originating in the diffusion is of order $\varepsilon$. This case can be seen in Figure \ref{fig:HighDeltaDifferentNoise} a). The DAP exerts a force inducing a constant velocity (see Section \ref{sec:4b}) resulting in parabola in a squared displacement plot. The corresponding depot energy can be seen in Figure \ref{fig:HighDeltaDifferentNoise} b), and a comparison to Figure \ref{fig:HighDeltaDifferentNoise} a) shows that the parabolic behaviour persists as long as the DAP drains its depot energy.

If we assume stronger noise the stochastic analysis in the $(\varphi,I)$-space becomes more complicated. In the presence of fixed points $\varphi^\ast$, trajectories spend a transient time in their vicinity, before looping around to $\varphi^\ast\pm2\pi$ and repeating the same process. The onset of this effect can be seen in Figure \ref{fig:HighDeltaDifferentNoise} c), where transient times are spent near $\varphi^\ast+2\pi n$, for integer $n$.

\section{Conclusion} \label{sec:6}
Over the course of this paper we have shown how a dynamical active particle converts its internal energy in the overdamped limit. The surprising result is that the overdamped dynamical active particle has the same efficiency as in the underdamped case. This is the case for the two identified parameter regimes, with completely different dynamics.

In section \ref{sec:2} we briefly reviewed the DAP and its key features, including the down-conversion from high frequency degrees of freedom to low frequency degrees of freedom over a long time scale. In section \ref{sec:3} we derived the overdamped equations from the four initial Hamiltonian equations and further reduced the number of equations to two. We presented a numerical integration of the overdamped equations of motion, which showed emptying of the depot degree of freedom for two completely different dynamical regimes of small and large travel distances. Both of the regimes resulted in a displacement of the dynamical active particle.

In section \ref{sec:4} we analysed the two opposing parameter regimes. In the first regime we showed by using the averaging theorem and an expansion in a small parameter that the mean change of the depot energy follows a simple solvable differential equation. In the second regime we built our analysis around fixed points in the phase. There we showed that the energy depot drains linearly and results in the same critical speed known from the underdamped equations.

We compared the two overdamped evolutions to an instantaneous energy transfer. We showed that the displacement generated by the dynamical active particle is finite, even in the overdamped limit, while for the instantaneous energy transfer, the displacement vanished in the overdamped limit. In section \ref{sec:5} we built on the analytic approximations and discussed the mean squared displacement for appropriate noise intensities in both of the regimes.

The main result of this paper is that the dynamical active particle does work in an overdamped environment. It not only functions but it does so with the same efficiency as in the underdamped case. We must furthermore keep in mind that in the underdamped case only a fraction of trajectories show an energy exchange. This fraction is of order $\sqrt{\varepsilon}$, so it might be quite small. This means that the dynamical active particle is actually \emph{more efficient} in the overdamped regime because almost every initial condition leads to active motion powered by the depot, independent of $\varepsilon$. 

The dynamical active particle does not simply survive the exposure to an overdamped environment: it flourishes in it.

\begin{acknowledgments}
The author thanks James R. Anglin for the valuable discussion on the topic and acknowledges the support from State Research Center OPTIMAS and the Deutsche Forschungsgemeinschaft (DFG) through SFB/TR185 (OSCAR), Project No. 277625399.
\end{acknowledgments}

\bibliography{citation}

\end{document}